\documentstyle[aps,graphics,multicol,psfig,epsfig,preprint,color]{revtex}

\begin{document}

\begin{center}

{\large {\bf From polymers to proteins -- novel phases of short  compact
tubes} }\\

\vskip 2.0cm {\sc Jayanth R.  Banavar$^{\dag}$ \& Amos Maritan$^{\ddag}$
}

\vskip 1.0cm

\end{center}

$^{\dag}$  Department of Physics, 104 Davey Laboratory, The Pennsylvania
State University, University Park, Pennsylvania 16802 \\

$^{\ddag}$  International School for Advanced Studies  (S.I.S.S.A.), Via
Beirut 2-4, 34014 Trieste, INFM and the Abdus Salam International Center
for Theoretical Physics, Trieste, Italy \\

\vskip 2.0cm

{\bf A framework is presented for understanding  the common character of
proteins.  Proteins are linear chain  molecules -- however, the simple 
model of a polymer viewed as  spheres  tethered  together  does not  account for
many of the observed characteristics of protein structures.  
We show here that the notion of a tube
of  non-zero  thickness  allows one to bridge the  conventional  compact
polymer   phase  with  a  novel  phase   employed  by  Nature  to  house
biomolecular  structures.   We build on the idea that a non-singular
continuum description of a tube (or a sheet) of arbitrary thickness
entails discarding pairwise interactions and using appropriately chosen many
body interactions.  We  suggest  that the  structures  of  folded
proteins are selected based on geometrical considerations and are poised
at the edge of compaction,   thus accounting  for  their   versatility  and
flexibility.  We present an  explanation  for why helices and sheets are
the building blocks of protein structures.  } 

\newpage
\tableofcontents
\newpage

\section{Introduction}
Recent years have  witnessed  gigantic  leaps in the field of  molecular
biology culminating in the sequencing of the human genome as reported in two
historic issues of Science (Volume 291, issue 5507) and Nature (Volume
409, issue 6822) in 2001.  
Base pairing and the  remarkable  structure of the DNA molecule  (Watson
and  Crick,  1953)  provide  a  very  efficient  means  of  storing  and
replicating  genetic information.  The  principal  role of genes is to
serve as a template for the synthesis of m-RNAs that, in turn, are 
``translated" by ribosomes into the polypeptide chains which then fold into active 
proteins.
These  proteins  are the work  horse  molecules  of life.  They not only
carry out a dizzying array of functions but also they interact with each
other and play a role in  turning  the genes on or off.  There is little
variability in the structure of the information  carrying molecule, DNA.
On the other hand, there are several  thousand  geometries  which folded
proteins can adopt and these structures  determine the  functionality of
the proteins (Branden and Tooze, 1999; Creighton, 1993; Fersht, 1998).  \\

Proteins  are the basic  constituents of all living cells.
Some familiar examples of proteins are hemoglobin,  which delivers oxygen
to our tissues,  actin and myosin  which facilitate the contraction of
our muscles,
insulin  which is secreted in the pancreas and signals our body to store
excess sugar and antibodies  that fight  infection.  Marvelous  machines
within the cell known as ribosomes  make proteins by stringing  together
little  chemical  entities  called amino acids into long linear  chains.
There are twenty  types of amino acids which  differ  only in their side
chains.  The  protein  backbone  as well as some of the side  chains are
hydrophobic  and shy away from water,  while other side chains are polar
and yet others have charges associated with them.  \\

Our focus is on small,  globular  proteins,  which, under  physiological
conditions,  fold  rapidly  and  reproducibly   (Anfinsen,  1973)  in  a
cooperative  fashion into a somewhat compact state in order to expel the
water from the core of the folded structure, which predominantly  houses
the  hydrophobic  amino  acids.  Thus there is an  effective  attraction
between the hydrophobic  amino acids arising from their shared  tendency
to avoid the water.  \\

For proteins, form determines function.  The structure of the protein in
its folded state (also called its native state  structure)  controls its
functionality  
(Branden and Tooze, 1999; Creighton, 1993; Fersht, 1998).
The rich variety of amino acids allows
for many  sequences to have the same native state  structure.  Thus even
though  our  human  body may have  more  than  100,000  proteins,  it is
believed  that the number of  distinct  folds  that they  adopt in their
native  state,  is  only  a  few  thousand  in  all   (Chothia,   1992).
Furthermore,  these  folds are  beautiful  (Levitt  and  Chothia,  1976;
Chothia,  1984) -- they are not just any compact  form but,  rather, are
made up of building blocks of helices and sheet-like  planar  structures
with tight loops connecting these secondary motifs (see Figure 1).  \\

In 1939, J.  D.  Bernal (1939) stated the challenge  associated with the
protein problem:  {\em Any effective  picture of protein  structure must
provide at the same time for the common  character  of all  proteins  as
exemplified  by their many chemical and physical  similarities,  and for
the highly specific nature of each protein type.}  Despite many advances
in  experiments  on proteins and the advent of powerful  computers,  the
problem  has  remained  largely  unsolved.  The  key  components  of the
problem are protein  folding and design:  protein  folding  entails  the
prediction  of the folded  geometry of a protein  given its  sequence of
amino acids while the design  problem  involves  the  prediction  of the
amino acid sequence which would fold into a putative  target  structure.
It is  probably  not too  surprising  that  progress  has been  somewhat
limited  because,  until  now,  there  has not been  any  simple unifying 
framework  for
understanding  the common  character of all proteins.  The principal aim
of this  colloquium  is to address  this issue.  Such a  framework  must
provide an explanation for the relatively small number of protein native
structures,  for why the  building  blocks  of  protein  structures  are
helices and sheets,  for the highly  cooperative  nature of the  folding
transition  of  small  globular  proteins  and for the  versatility  and
flexibility  of protein  structures,  which account for the ability of the
proteins to
perform a wide range of functions.  \\

\section{Quantum chemistry scores a major success}

Linus Pauling and his  collaborators  (Pauling, Corey and Branson, 1951;
Pauling and Corey, 1951) invoked the  chemistry of covalent and hydrogen
bonds to show that  helices  and  sheets  were  periodically  repeatable
structures for which  appropriately  placed hydrogen bonds could provide
the scaffolding.  This stunning prediction was experimentally  confirmed
in short  order.  Unfortunately,  these  observations  do not  provide a
complete  explanation  of the selection of the protein folds.  The  difficulty
arises  because  hydrogen  bonds can  equally  easily  form  between the
protein molecule and the water surrounding it.  While helices and sheets
are nicely  stabilized by hydrogen bonds, one may construct other viable
structures  which do not have helices or sheets as the  building  blocks
but yet have a large  number of  hydrogen  bonds and  hence a  favorable
energy.  \\

A protein is complex because of the many features that one is confronted
with.  As mentioned  before, we need to deal with twenty  types of amino
acids with their  individual  properties and, in addition,  there is the
crucial role played by the solvent.  A first  principles  approach might
consist of  considering  all the numerous  atoms making up a protein and
the  surrounding  solvent and carrying  out some really  heavy  computer
calculations to simulate the folding process.  Very quickly one realizes
that, with the somewhat imperfect  knowledge of the interactions and the
sheer  magnitude of the job at hand, this  approach is not too likely to
yield  qualitatively  new  insights  into the protein  folding  problem.
Furthermore,  one might worry  that, at best, one would be able to mimic
Nature but would one obtain an understanding of Nature?  \\

\section{A  physics  approach leads to a disconnect  between the compact
polymer phase and the novel phase adopted by protein structures}

Let us now consider the protein  problem  afresh from a physics point of
view and attempt to identify  the key issues.  It is of course  possible
and, one might fear, even likely that many of the details are crucial in
understanding  the  intricate  behavior  of  proteins.  In order to make
progress, we will take the approach, though, of looking at what we might
imagine to be the the most  essential  features and adding in details as
we require  them.  This will allow us to retain  some  control  over our
understanding  and we will be able to assess, a posteriori, the relative
importance of the features that we may have to incorporate.  \\

The approach is analogous to one commonly  used in physics  (Chaikin and
Lubensky,  1995) of distilling out just the most essential  features for
understanding  emergent  phenomena.  For  example,  one can use  general
geometrical and symmetry  arguments to predict the different  classes of
crystal  structures.  The existence of these structures does not rely on
quantum  mechanics or on chemistry.  They are a consequence  of a deeper
and more general  mathematical  framework.  Of course,  given a chemical
compound like common salt, a careful quantum mechanical study would show
that sodium chloride  adopts the face centered cubic lattice  structure.
Also, a clever  grocer  would  use the same  crystal  structure  for the
efficient   packing  of  fruits.  Thus  the  structures   transcend  the
specifics of the chemical  entities  that are housed in them.  One might
therefore  seek to  determine  what  the  analogous  structures  are for
protein  native  states  that  are  determined   merely  by  geometrical
considerations.  What are the bare  essentials  that determine the novel
phase adopted by biopolymers such as proteins?  \\

Proteins are linear  chains and  ignoring  the details of the amino acid
side-chains,  all proteins have a backbone.  A protein folds  because of
hydrophobicity  or the tendency of certain  amino acids to shy away from
water.  In the  folded  state,  therefore,  one  would  like  to  have a
conformation  which  squeezes the water out from certain  regions of the
protein populated by the hydrophobic amino acids.  As stated before, the
simplest way of encapsulating such a tendency for compaction is by means
of an  effective  attractive  interaction  between  the  backbone  atoms
promoting a somewhat compact native state.  \\

An early success of the physics  approach was the work of  Ramachandran,
Ramakrishnan and Sasisekharan  (Ramachandran and Sasisekharan, 1968), as
embodied in the  Ramachandran  plot.  They showed that steric  constraints,
relating to or involving the arrangement of atoms in space, 
alone  dictated  that  the  backbone  conformations  of  a  protein  lie
predominantly  in two  regions  of the space of the so called  torsional
angles  corresponding  
to $\alpha$-helical  and  $\beta$-strand  conformations
(see Figures 1-3).  
In other
words, the high cost associated  with the overlap of two atoms viewed as
hard spheres leads to conformations  which are consistent with the local
structure associated with a helix or a sheet.  \\

We hit a snag in our thought experiment -- careful computational studies
(Hunt et al.,  1994; Yee et al.,  1994)  have  shown  that the  standard
polymer model of chain molecules,  viewed as spheres tethered  together,
when  subjected  to  interactions   which  promote   compactness,   have
innumerable conformations almost none of which have any secondary motifs.
In contrast, proteins have a limited  
number of folds to choose from for their native
state  structure  and  the  energy   landscape  is  vastly  simpler.  In
addition,  the  structures  in  the  polymer  phase  are  not  specially
sensitive to  perturbations  and are thus not as flexible and  versatile as
protein native state structures are in order to accommodate the dizzying
array of  functions  that  proteins  perform.  Indeed,  there  has  been
somewhat of a disconnect  between the familiar compact polymer phase and
the novel phase used by Nature to house  biomolecules.  To quote from P.
J.  Flory  (1969):  {\em  Synthetic  analogs of  globular  proteins  are
unknown.  The  capability  of  adopting a dense  globular  configuration
stabilized by  self-interactions  and of transforming  reversibly to the
random coil are  peculiar to the chain  molecules  of globular  proteins
alone.}  \\

\section{Protein backbone viewed as a tube}

So what new feature should we incorporate  next?  Are the details of the
amino  acids  important?  We expect not  because  it is known  that many
sequences fold into the same native state structure  
(Branden and Tooze, 1999; Creighton, 1993; Fersht, 1998).  At
a  somewhat   simpler   level,  we  recall  the  work  of   Ramachandran
(Ramachandran   and   Sasisekharan,   1968),  who  showed   that  steric
interactions  or the  undesirability  of two atoms to sit on top of each
other, even when the atoms are treated as  effective  hard  spheres, 
lead to  certain  regions of
conformational  space being  excluded for a protein  chain (Rose, 1996).
The side  chains of the  amino  acids  occupy  space as well, and thus it
seems  important  to allow for room around the  backbone to  accommodate
these atoms.  So we proceed by  incorporating a new ingredient -- let us
treat the  protein  backbone  not as a chain of spheres but as a tube of
non-zero thickness  analogous to a garden hose.  So how does such a tube
behave if it has an effective attractive  self-interaction  that tends to make its
conformation  somewhat  compact?  There  is some  hope  on the  horizon,
because the  problem is  enriched  and we now have two length  scales to
play  with, the  thickness  of the tube and the range of the  attractive
interactions.  \\

It is useful to consider what the standard model of a chain  represented
as tethered spheres is missing.  For  unconstrained  particles,  spheres
are the simplest  objects that one might consider.  Of course,  symmetry
matters a great deal and when these  spheres are replaced by  asymmetric
objects, one gets a host of qualitatively new liquid crystalline phases
(Chaikin and Lubensky, 1995).
There  are  two  simple  ingredients  associated  with  a  chain  -- the
particles  are  tethered to each other  (which is  well-captured  by the
standard model of tethered spheres) and associated with each particle of
the  chain is a  special  direction  representing  the  local  direction
associated  with the chain as defined by the adjacent  particles at that
location.  This selection of a local direction  immediately leads to the
requirement that the symmetrical  spherical objects comprising the chain
must be replaced by anisotropic objects (such as coins) for which one of
the three directions is different from the other two.  Thus, if one were
to think of a chain as being made up of stacked coins  instead of  spheres,  one
naturally  arrives at a picture  of a tube.  Indeed,  previous  analysis
(Banavar et al., 2002 (a); Banavar et al., 2002 (b)) of the native state
structures of proteins has shown that a protein  backbone may be thought
of  approximately  as a uniform  tube of  radius  $2.7  \AA$.  Before we
explore the phases associated with a tube subject to compaction, we will
have an interlude where we will revisit some issues in polymer  physics.
\\

\section{Strings, sheets and many-body interactions}

Strings  and  chains  have been  studied  over the years in the field of
polymer  physics.  Tubes of non-zero  thickness are ubiquitous -- garden
hoses  and   spaghetti  populate  our  houses.  How does one
mathematically  describe a tube of non-zero  thickness in the  continuum
limit?  Alas, a visit to the  library  confirms  our worst  fears,  this
elementary problem has not been tackled before.  A continuum description
of a string was put forward by Sam Edwards (Doi and Edwards, 1993) -- it
captures  self-avoidance by means of a singular delta function repulsion
between  different  parts of a string.  The delta  function  describes a
situation  in which  the  repulsive  interaction  is  infinitely  strong
precisely  when  there is an exact  overlap  and  zero  otherwise.  This
description is therefore valid only for an infinitesimally  thin string.
An associated  complication  is that the analysis of a continuum  string
requires  the use of  renormalization  group  theory to  regularize  the
theory by introducing a lower-length scale cut-off combined with a proof
that the behavior, at long length scales, is independent of this cut-off length
scale.  
Unfortunately,  the  renormalization  group theory  analysis, in
this context, is peripheral to the physics
being studied.  \\

Recently, with the help of two  mathematicians,  Oscar Gonzalez and John
Maddocks, we were able to write down a singularity-free  description of
manifolds  such as chains  or  sheets  (Banavar  et al.,  2002(c)).  The
solution is very simple but not  intuitively  obvious.  In science,  the
starting  point  for  describing  interacting  matter  is  by  means  of
pair-wise  interactions.  In order to describe  your  interactions  with
your friends, it is a good  starting  point to consider  your  pair-wise
interactions  with  each  of  them  --  your  true  interaction  will be
different from this only because of genuine many body interactions  that
may  be  thought  of as  higher  order  corrections.  With  a  pair-wise
interaction,  there is only one length scale that one can construct from
a  knowledge  of where you are and where your  friend  is.  This  length
scale is your mutual  distance.  One can define  potential  energies  of
interaction  between you and your  friend  which  depends on this length
scale.  Generically,  such an interaction may be one in which if you and
your friend are separated by a  sufficiently  long  distance, you do not
talk to each  other and  there is no  interaction.  There is an  optimal
distance  between you and your friend  where the  interaction  is at its
happiest.  Any  closer  approach  leads  to a  higher  energy  with  the
potential energy becoming  infinitely  large when you sit on top of each
other.  \\

Unfortunately,  such an analysis is not very  helpful  when you and your
friends (and your  enemies)  are formed into a conga line by someone who
does not know what your personal  relationships are.  Let us assume that
one is working again with pair-wise  interactions  and you are told that
two people are spatially  (not  necessarily  emotionally)  close to each
other.  With  that  information  alone,  you  will  not be able  to tell
whether  the two people  are from  different  parts of the chain and are
close to each other  because  they like each  other or  whether  the two
people are sworn  enemies  who happen to be close to each  other  simply
because  they were  constrained  to be next to each  other in the  conga
line.  In other words, pair-wise  interactions merely provide the mutual
distance but not the context in which the interacting  particles  exist.
\\

The basic idea behind the development of a continuum theory of a tube of
non-zero  thickness is to discard  pair-wise  interactions  and consider
appropriately  chosen three-body  interactions as the basic  interacting
unit.  The requirements for a well-founded  theory are that one ought to
be  able  to  take a  continuum  limit  on  increasing  the  density  of
particles,  that  self-interactions  be properly  taken into account and
that there be a characteristic microscopic length other than the spacing
between neighboring particles along the string.  \\

Let us consider a three-body  potential  characterizing  the interaction
between  three  particles,  which lie on the corners of a triangle.  Let
the sides of the triangle  have  magnitudes  $r_1$, $r_2$ and $r_3$.  In
order to specify a triangle  uniquely, one needs three  attributes.  The
potential  of  interaction  can  therefore  depend on three  independent
length  scales,  which are  invariant  under  translation,  rotation and
permutation of the three particles.  We choose these length scales to be
the perimeter,  $P$, of the triangle, the ratio of the area, $A$, of the
triangle  to its  perimeter,  $P$, and  finally  $r_1 r_2 r_3 / A$.  The
first two  lengths do not cure the  problems  alluded  to before -- they
both  vanish  when  the  particles   approach   each  other  and  cannot
distinguish  between particles from the same region or different regions
of the  string.  The  third  length  scale is  proportional  to $R$, the
radius of a circle drawn through the three  particles  and has proved to
be valuable for the study of knots (Gonzalez and Maddocks,  1999).  This
length scale neatly solves the contextual problem mentioned above.  When
two parts of a chain  come  together,  the  radius  of a circle  passing
through two of the  particles  on one side of the chain and one particle
from the  other  side of the  chain  turns  out to be a  measure  of the
distance of approach of the two sides of the chain.  On the other  hand,
when one considers three  particles  consecutively  along the chain, the
radius of the circle passing  through them is simply the local radius of
curvature.  Indeed when three such particles  form a straight  line, the
radius  goes to  infinity  and the three  particles  essentially  become
non-interacting.  The straight line  configuration  is the best that the
particles  can do in terms of staying  away from each  other  given that
they are constrained to be neighbors along the chain.  Our suggestion is
to use a generic  potential  energy  function  such as the one described
previously but with this three-body radius as its argument.  \\

How might one define the  thickness  of a tube  associated  with a chain
configuration?  A simple  procedure  would be to  construct a tube whose
axis coincides  with the chain and inflate the tube  uniformly  until it
intersects  with itself or has sharp  corners.  A natural  definition of
the  thickness is then the radius of this  largest  tube (Katritch et al., 1996).  
A tube with a
large thickness has more space for internal  rearrangements  of the side
chains of the amino acids than a thinner tube.  This  thickness can also
be obtained using the  three-body  interactions  by computing the radius
associated with all triplets (contiguous or otherwise) and selecting the
smallest among these radii (Gonzalez and Maddocks,  1999).  A simple way
of describing a tube of non-zero  thickness in the continuum limit is to
discard  pair-wise  interactions and to consider  triplet  interactions.
One may choose a simple potential  energy which is a sum of 3-body terms
whose  argument is the 3-body  radius and whose form has a hard-core  at
short  distances -- any radius (local or  non-local)  is forbidden  from
taking on a value less than the  thickness  of the tube (see Figure 2).  
Likewise,  one
may write down a continuum description for the self-avoidance of a sheet
of paper (such as the page you are  reading) of  non-zero  thickness  by
discarding   pairwise  and   three-body   interactions   and   employing
appropriately  chosen 4-body  interactions as the basic interacting unit
(Banavar et al., 2002(c)).  \\

The new insight that one obtains with this continuum  description is the
important role of three-body  interactions  in  characterizing  tubes of
non-zero  thickness.  It is important to stress that this elimination of
pair-wise  potentials  and their  replacement  by  effective  three-body
potentials  is  necessary  only  in  the  continuum   limit.  Also,  the
potentials  we are  discussing  are  effective  potentials  obtained  on
integrating what one hopes are irrelevant finer degrees of freedom.  
Our formulation not only
allows one to carry out a continuum
study of thick polymer chains but also it is useful for the study of
chains in a given knot class or with a  fixed number of knots. Any model
employing a pairwise potential allows self-intersections, albeit with
an energy penalty, so that the topology of the polymer chain (as
measured by the knotting number or linking number) can be changed at will. 
This, of course, does not happen in real life with closed chains. Thus
our non-singular many body potential allows one to formulate an analytic
attack on the entropic exponents and weights of polymer configurations with
a fixed linking number.
\\

\section{Marginally compact tubes }

We return  now to the  protein  backbone  viewed  as a tube of  non-zero
thickness.  Consider a uniform  tube  undergoing
compaction  to expel the water  away  from the  interior  of its stucture
in the  folded
state.  (We alert the reader to the fact 
that the tube we are considering is not 
hollow.)  The backbone of all amino acids contains a carbon 
atom which is
called a $C_{\alpha}$ atom. In a coarse-grained description, this atom 
may be chosen as the representative of the amino acid.  For specificity, let us 
consider a  discrete  chain  of
$C_{\alpha}$  atoms of the  protein  backbone.
As we have discussed, the notion of a tube
thickness is captured by ensuring that none of the  three-body  radii is
smaller than a threshold  value equal to the radius of the tube.  Let us
also postulate that the attractive  interactions promoting compaction are
pairwise and have a given range.  (Because we are  considering a discrete
situation, it is quite valid to have  pairwise  interactions.)  There is
one dimensionless quantity, which we will call $X$, that we will need to
specify, which is the ratio of the thickness of the tube to the range of
the attractive interactions.  \\

When $X$ is very  large  compared  to 1, the  tube is so fat  that it is
unable to benefit from the attractive  interactions.  The constraints of
the three body interaction  dominate (the pairwise  interaction plays no
role)  and one then  obtains  a  swollen  phase  which  consists  of all
self-avoiding   conformations   which  satisfy  the  three-body   radius
constraint  associated  with the non-zero tube  thickness.   
A vast majority
of these conformations are ineffective in expelling the water from the
interior of the structure.   The non-zero
thickness  is loosely  analogous  to restricted  space that  others are not
allowed  to  trespass.  Imagine  that your  friend  sits in the
center  of his/her  room  and  requests  that no one  enter  the  room.  The
thickness then is  proportional  to the width of the room.  If the range
of  attractive  interactions  is very small  compared to this size, your
ability to benefit from interactions with him/her is compromised by the fact
that you cannot enter the room and for all practical  purposes, it is as
though  your  interactions  with  him/her  were  turned  off.  At the  other
extreme,  for a tube  with a very  small  $X$  compared  to 1, one  also
obtains  many, many  conformations.  This  is because, in the room  analogy,
your  interaction  with your friend is  sufficiently  long range so that
there is a lot of  flexibility  in where you position  yourself.  From a
dynamical  point of view,  the  structures  obtained when
$X << 1$ are  somewhat inaccessible 
because  the energy  landscape  is  studded  with  numerous
multiple minima.  This situation is one in which the pairwise attractive
interactions  dominate  and the  three  body  radii  constraints  do not
matter.  \\

On varying  $X$, we  therefore  expect two  regimes,  the phase  with an
effective  long  range  attraction  and the  swollen  phase,  both  with
tremendous  degeneracies.  There is a twilight  zone  between  these two
phases,  viewed as day and night,   when $X$ is
just shy of 1 (Figure 4).  
(We alert  experts in the protein  field that this crossover 
which we characterize  colloquially as a twilight zone has no
relationship  to and should not be  confused  with the same  terminology
sometimes used in the studies of sequence similarity.)  
In this twilight
zone, there is a rich interplay of the pairwise attractive  interactions
and the  constraints  imposed by the three-body  interaction.  This is a
situation  in which you are able to  interact  with your  friend but can
only do so by positioning yourself  right outside his/her room.  \\

In the twilight zone, a tube is barely  able to
avail itself of the attractive  interactions promoting compaction.
In this region of parameter space,  the
forces  promoting  compaction just set in and one would expect to obtain
marginally compact  structures which have the ability to expel the water
from the  interior.  In addition,  because the scale of the  interaction
strength is relatively small, one would expect a low ordering transition
temperature with entropic effects not being too important.  
Furthermore, the physical  picture of a tube (recall that a tube 
can be thought of as many
anisotropic  coins tethered  together)  leads to a strongly  anisotropic
interaction  between  nearby tube  segments  -- it is better to position
them  parallel  rather than  perpendicular  to each other.  Thus, in the
twilight  zone, one has a  relatively  weak and strongly anisotropic
interaction. 
Because the tube segments have to position themselves next to each other
and with the  right  relative  orientation  in  order  to  avail  themselves of the
attractive  self-interaction, one would expect a cooperative  transition
with few  intermediates -- the tube will need to snap into its correctly
folded   configuration.  Also,  because  of  the  loss  of   flexibility
regarding  the  relative  positioning  and  orientation  of nearby  tube
segments, one would expect a large decrease in the degeneracy.  \\

In proteins, why is this effective  value of $X$ tuned to be so close to
its  transition  value  of 1?  The  answer  lies in the  fact  that  the
atomistic  scale  interactions  are short range due to the  
screening effects of the water
and, at the  microscopic
scale, the squeezing out of water is  facilitated  by the outer atoms of
nearby  side  chains  coming  together.  In a  coarse-grained  level of
description, this translates  into a value of $X$
which is  close  to  1,  because,  on the one  hand,  the
necessity  of having some wiggle  room for the side  chains of the amino
acids leads to the tube picture and determines the tube thickness and, on
the other hand,  the same  side-chains  are  responsible  for 
and control the range of the  attractive
interactions promoting compaction.  \\

There are several  significant  advantages in the system being poised in
this twilight zone  and having a limited number
of  marginally   compact   structures  as  the  candidate  native  state
conformations.  In the thermodynamic limit of a tube of infinite length,
there is a first order transition, on decreasing the tube thickness, between
a swollen phase and a compact phase.  This phase transition is characterized,
nevertheless, by a diverging length scale -- the propensity for nearby tube
segments to be aligned just right with respect to each other leads to
a diverging persistence length, defined as the characteristic length over
which memory of the tube orientation is preserved. \\

Let us  briefly review the well studied  subject of
phase transitions and critical  phenomena  (Stanley, 1999).  Examples of
critical   points   include  a  magnet  at  the  onset  of  ordering,  a
liquid-vapor  system at the critical  temperature  and pressure, 
and a binary  liquid  system which is about to phase  separate.  The key
point is that the  fluctuations  in a system at its critical point occur
at  all  scales  and  the  system  is  exquisitely   sensitive  to  tiny
perturbations.  Even though  sharp phase  transitions  can occur only in
infinitely large systems, behavior akin to that at a phase transition is
observed  for finite size  systems as well.  Indeed, for a system near a
critical  point, the  largest  scale over  which  fluctuations  occur is
determined  either by how far away one is from the critical  point or by
the finite size of the system.  \\

A magnet at low  temperatures  compared to its critical  temperature  is
well  magnetized  and is not very  sensitive to a tiny  external  field.
After all, when the  magnetization is large, small  perturbations do not
lead  to  major   consequences.  Similarly,   a  magnet   at  very  high
temperatures  is not very sensitive to a tiny external field because the
strong thermal  fluctuations  dominate and the ordering  tendencies  are
rather  small.  However, at the critical  point, where there is about to
be an onset of the magnetization, the system is very sensitive to
an applied magnetic field and indeed the magnetic  susceptibility for an
infinite system diverges.  \\

Nature, in her desire to design proteins to serve as smart and versatile
machines, has used a system  poised near a phase  transition  to exploit
this   sensitivity.  Indeed,  it  is  well  known  that   proteins   
utilize 
conformational  flexibility (Jacobs et al., 2001)
to  achieve  optimal  catalytic  properties
(Branden and Tooze, 1999; Creighton, 1993; Fersht, 1998).
That  protein  structures  are  poised  near  a  phase
transition  provides the versatility and the flexibility  needed for the
amazing range of functions that proteins perform.  \\

In this  marginally  compact  state,  the  number of  candidate  protein
structures is somewhat limited.  An energy landscape with relatively few
energy minima associated with the protein folds has several consequences
and  advantages.  First,  each of these minima will have a  correspondingly
large  basin of  attraction.  Second,  a  protein  sequence  has  only a
limited menu to choose from in deciding  what its native  state ought to
be.  A simple analogy is the greater  ease we have in a  selecting  from a
restaurant menu containing a few items than from one with innumerable 
choices. This  selection  is further  reduced  by the  requirement  that the
native state be compatible with general chemical  affinities and be free
of steric clashes.  The quality of match (Banavar and Maritan, 2001)
between a  sequence and a putative 
native state structure can be assessed by considering the propensity of the individual
amino acids to be in distinct secondary structure elements 
such as the $\alpha$-helix or 
a $\beta$-sheet, their likelihood of being buried or exposed and the
degree to which the native state structure accommodates the `desire' for certain
pairs of amino acids to be in the vicinity of or away from each other.
Strikingly, as seen in protein  engineering  experiments
(Fersht,  1998), the ultimate  choice of which fold a sequence adopts
is dictated
by  a  small  number  of  key  amino acids which  have  distinctly  better
environments  in the native  state  than in  competing  folds.   The
limited number of these special folds  underscores  the key role played by the
native state topology in  determining  many of the essential  aspects of
protein  folding  (Micheletti  et al., 1999;  Maritan  et al., 2000 (a);
Maritan  et  al.,  2000(b)  ;  Baker,  2000).  The  powerful  forces  of
evolution  (Lesk and Chothia, 1980) operate within the fixed  playground
of these selected folds  yielding  better or more  versatile  sequences.
Indeed, multiple protein functionalities can arise within the context of
a single fold (Holm and Sander, 1997).  \\

\section{Building blocks of protein structures}

In order to determine the nature of the twilight zone structures  
(Banavar et al., 2002(a); 2002(b)) that a
protein  would adopt in its native  state and be  efficient  in terms of
expelling water from its interior, we will begin by considering a coarse-grained
representation  of the protein as a uniform  tube of, say, unit
radius  (recall  that  1  unit  is  approximately  $2.7  \AA$).  In  the
discretized case, one may consider the backbone of the protein with just
the  $C_{\alpha}$  atoms.  The tube  thickness  at a given  $C_{\alpha}$
location is obtained by considering all triplets of  $C_{\alpha}$  atoms
including the  $C_{\alpha}$  at that location and selecting the smallest
among  all  their  radii.  Recall  the  notion  of  the  private   space
associated with the thickness,  which requires that no three body radius
be smaller  than 1 unit.  In order to  promote  conformations  which are
efficient in squeezing the water from the interior of the  structure, we
could invoke an effective  potential which promotes radii close to unity
or the tube thickness.  \\

Let  us  ask  now  what  one  obtains  for  the  energetically   favored
conformations  of a short  chain  made  up of  discrete  particles  with
three-body  potentials whose energy is lowest when the radius is 1 unit.
The simplest  starting point for obtaining a happy  configuration  is to
choose the local radius of curvature (the radius  associated  with three
contiguous  particles)  to be 1 unit.  Winding the chain around a circle
will  lead  to  an  overlap  of  the  chain  with  itself  and  that  is
prohibitively  expensive.  So one would  instead  choose a helix  with a
local radius of curvature equal to 1 unit.  But how would one select the
pitch of the  optimal  helix?  The  pitch  would be  chosen  so that the
radius  characterizing  the three body interaction  comprising a pair of
particles  from one turn of the helix and another  from the next turn is
again  equal to unity,  thus  lowering  the  energy.  This  picks  out a
special   pitch  to  radius   ratio  of  the   helix.  Strikingly,   the
corresponding  ratio in helices of proteins  is within a few  percent of
this prediction  (Maritan et al., 2000(b),  Stasiak and Maddocks, 2000).
Furthermore, the tube segments corresponding to neighboring turns of the
helix are  oriented  parallel to each other and respect  the  anisotropy
inherent  in a  tube-like  description  (see  Figure  2).  This  helical
conformation corresponds to the space-filling  configuration of a garden
hose in which the local radius of  curvature  equals the tube  thickness
(any smaller local radius is disallowed) and in which  successive  turns
of the hose lie on top of each other with no intervening space.  \\

How would one deal with a situation  when the bulky side chains of amino
acids do not allow a segment of a chain to be placed in a tight  turn of
such a small  radius?  The local  radius of  curvature  would have to be
larger than $2.7 \AA$.  In this situation, an alternative way to promote
triplets  having a radius of 1 unit (or  around  $2.7  \AA$) is  through
non-local interactions.  One possibility that would be cumbersome from a
folding  point of view is to have  multiple  helices with a larger local
radius of curvature winding around each other.  \\

A more versatile way to obtain  non-local  interactions is by means of a
sheet.  First,  a  strand  in an  extended  conformation  would  form to
locally accommodate the larger radius of curvature enforced by the local
steric  incompatibility.  In  order  to have  as  many  triplets  of the
desired radius of 1 unit as possible, one would need interactions with a
different part of the chain and the problem  reduces to determining  the
optimal  placement of two  essentially  independent  parts of the chain.
From  symmetry  considerations,  one  would  expect  the most  favorable
circumstances  to occur when two such identical  strands from  different
parts of the chain are in the  vicinity of each other and when they both
lie  essentially  in a  plane.  The  three-body  interaction  encourages
planarity by not only  allowing for a harmonious  fit of the strands but
also  providing  room for the side  chains  perpendicular  to the plane.
Pauling  and Corey  (1951) had shown that two  neighboring  strands in a
protein  are  replicas  of each  other or mirror  images  of each  other
(Richardson,  1997)  (corresponding to parallel and antiparallel  sheets
respectively) in terms of the backbone  atoms.  They are located at an
optimal  distance  from each  other,  which  allows  the  formation  of a
supporting  framework for the assembly of the strands  based on hydrogen
bonds  between  atoms in  neighboring  strands.  A sheet is formed  by a
repetition  of the  same  process  (Figure  3).  In this  case as  well,
adjoining  segments of the tube  (neighboring  strands) are  parallel to
each  other.  Strikingly,  one can show  analytically  (Banavar  et al.,
2000(b)) that the familiar  zig-zag pattern of the strands arise because
of the discrete  nature of the chain made up of the $C_\alpha$  atoms of
the backbone.  \\

When  one  considers   longer  segments  of  the  proteins,  it  is  not
energetically  favorable  to have just one  helix or one  sheet  because
distant regions would not necessarily have triplets characterized by the
preferred radius.  Thus there is a persistence  length associated with a
given secondary structure.  In order to assemble the tertiary structure,
which  provides  more  energetically   favored  triplets  from  distinct
secondary  structure  elements, one would need a mechanism for producing
tight  turns,  which  would  entail  having  a  small  local  radius  of
curvature.  This is facilitated by small amino acids like Glycine, which
is often found in backward  bends.  In reality,  therefore, a protein is
strictly speaking not characterized by a uniform thickness.  \\

The thickness  associated  with a tightly wound helical geometry  would be
expected to be slightly  less than that  associated  with a hairpin or a
sheet geometry.  But are there other  structures that might emerge which
have many triplets  having the optimal  radius?  A possibility  that one
might  expect is a saddle  structure  instead of a hairpin.  The easiest
way to visualize a saddle is to start with a planar  hairpin and bend it
into a three dimensional  object.  The distinct  advantage of doing this
is the ability to create additional contacts at the cost of reducing the
thickness  somewhat.  However,  Nature  does  not  seem  to  adopt  this
conformation in proteins because of the inability to form hydrogen bonds
and provide the necessary  scaffolding.  Nevertheless,  kissing hairpins
are found in RNA secondary structures (see Figure 5).\\

\section{Consequences of the tube picture}

Within the  hierarchical  picture of folding  (Baldwin  and Rose, 1999),
each  short  local  segment  of a  sequence  may be  associated  with  a
propensity  or ability to either  form very tight  turns (as in backward
bends), or the  regular  tight turn  associated  with a helix, or, indeed, a
desirability to be in a strand  conformation  with a larger local radius
of curvature.  This local  information  then has to be put together in a
global way in order to provide  stability  for the  strands by forming a
hairpin or a sheet  structure.  The  complexity  arises  because a short
segment of the sequence which is able to form a helix may instead choose
to form a strand in order to stabilize a nearby  segment  which can only
form a strand.  These decisions are of course non-local in character and
furthermore one has to ensure that all the turns can be made to assemble
the tertiary  structure  and that all the  hydrophobic  residues are shielded
from the water in the folded state.  \\

There are an  astronomical  number of sequences  that one can  construct,
even for modest  lengths.  Why then are  there so few  sequences  that are
protein-like?  More  generally, for purposes of protein  design, what is
the  selection  principle in sequence  space?  It is likely that, for an
overwhelming  majority of  sequences,  different  parts of the  sequence
would  attempt  to take on  conformations  corresponding  to  pieces  of
secondary  structure  that simply do not fit together to form one of the
folds.  This inherent  frustration is absent for protein-like  sequences
and is  responsible  for  selection in sequence  space  (Bryngelson  and
Wolynes, 1987).  The rich and varied  repertoire of amino acids has been
used by Nature in evolution  to design  sequences  that are able to fold
rapidly and  reproducibly  to just their native  states.  There are many
sequences that fold into a given  structure  because once a sequence has
selected  its  native  state   structure,  it  is  able  to  tolerate  a
significant degree of mutability except at certain key locations (Sander
and Schneider, 1991; Kamtekar et al., 1993; West et al., 1999).  
Also, such a design could be carried
out in order to create a folding 
funnel (Bryngelson et al., 1995; Dill and Chan, 1997)
with a  minimal amount of 
ruggedness in the energy landscape.  
\\

It is  interesting  to  consider  the  ground  state of many long  tubes
subject to  compaction.  
Packing  considerations  suggest that the tubes
become essentially  straight and parallel to each other and are arranged
(when viewed end on) in a triangular lattice, analogous to the Abrikosov
flux lattice phase in superconductors (Tinkham, 1996).  Returning to the
case of a single  tube, in the very long length  limit, a similar  phase
would be expected with the  additional  constraint of the bending of the
tube  segments at the ends.  One can show that, for a discrete  chain, a
planar  placement of zig-zag strands is able to accommodate  the largest
thickness  tube that can yet avail of the  attraction.   However,  the
thickness  for this  limiting  case is too  large to  produce  the 
three-dimensional  ordering  alluded  to  above.  It would be  interesting  to
consider  how  the  ground  state   structure   crosses  over  from  the
``flux-lattice"  type phase to the familiar  planar  phase.  Indeed, for
thick  tubes  of  moderate  length,  one  may  expect  to  form a  large
sheet-like structure analogous to the cross-$\beta$-scaffold observed as
a building block of amyloid fibrils  (Dobson,  1999; 2002).  Such fibrils have
been implicated in a variety of human  disorders  including  Alzheimer's
disease  and  spongiform   encephalopathies  such  as  Creutzfeldt-Jakob
disease.  
{\em The generic fibrillar forms of proteins can be regarded as the 
intrinsic `polymer' structure of a polypeptide chain } (Dobson, 2002)
and is a direct confirmation of the tube picture presented here.  \\

\section{Studies of short tubes}

We have carried out numerous
analytic and  computational  studies (Banavar et al., 2002(a); 
2002(b);  2002(d)) 
and have
quantitatively  confirmed the ideas presented here.  As an illustration,
Figure 5 shows the structures obtained in computer  simulations of short
tubes in the marginally  compact  phase (Banavar et al., 2002(b)).  
Helices and hairpins  (sheets)
are of course the well-known  building blocks of protein structures (see
Fig.  5 (A1) and (D1) for two examples from a protein and (A2), (D2) and
(D3) for the tube  structures  in our  simulations).  
It is  interesting  to
note that some of the  other  marginally  compact  conformations  bear a
qualitative  resemblance  to  secondary  folds in  biopolymers.  Helices
analogous to Fig.  5 (A3) with an irregular  contact map occur, e.g., in
the HMG protein NHP6a (Allain et al., 1999) with pdb code 1CG7.  Fig. 5 
(C1) shows the ``kissing hairpins'' (Chang and Tinoco, 1997) of RNA (pdb
code 1KIS), each of which is a distorted and twisted  hairpin  structure
while Fig.  5 (C2) is the corresponding tube conformation.  Fig.  5 (B1)
shows a helix of strands found  experimentally  in Zinc  metalloprotease
(Baumann  et al., 1993) (pdb code:  1KAP),  whereas  Fig.  5 (B2) is the
corresponding   marginally   compact   conformation   obtained   in  our
calculations.  \\

Specifically,  these studies have shown that a thick short tube in the twilight zone
assumes conformations corresponding to helices
of the correct  pitch to radius  ratio and  zig-zag 
hairpins  and sheets.   These  building
blocks of protein structures
are the only  ones
that are  effective  in
expelling  the water from their  interior.
Furthermore, these structures
are poised near a phase transition
of a new kind of the corresponding infinite-sized system.  \\

\section{Summary and Conclusions}

We have presented  a simple unifying framework for understanding the
common character of all proteins.  Our analysis is based on
just three ingredients -- all proteins share a backbone, 
there are effective forces which promote the folding of a protein
and the one and only new idea that a protein can be viewed as a tube.
We have shown how one may write
down a non-singular continuum description of a tube or a sheet of non-zero
thickness.  The recipe for doing this has the surprising feature that
pairwise interactions potentials need to be discarded and replaced by
appropriate many-body potentials. \\ 

We have considered  a situation in which
there is an attractive force, mimicking the hydrophobicity,  between
different parts of the tube.  New
physics arises from the interplay between two length scales: the
thickness of the tube and the range of attractive interactions.  
Many of the known polymer phases are found when the tube is very thin
compared to other length scales in the problem.  However, when the two
length scales become comparable, one obtains a novel phase of matter that is
used by proteins for its native state structures. This new phase
has many properties which explains the 
character of all small globular proteins, which do not depend on the
specific amino acid sequence.  These include
the ability of the folded structure to expel water efficiently 
from its interior,
the existence of a simple energy landscape with 
relatively few putative marginally compact
native state structures,
an explanation for many of the well-known characteristics of globular 
proteins such as  helices and hairpins and sheets 
being the building blocks of protein
structures,
the cooperative folding of small proteins,
generic formation of fibrils in tube-like polypeptide
chains  and
the acute sensitivity of protein structures to the
right types of perturbations thus accounting for their flexibility and
versatility.   \\

Many strategies for attacking the protein folding  problem have been put
forward which employ a coarse-grained  description (Banavar and Maritan,
2001).  None of the  currently  used  methods has been  successful.  Our
results suggest that a deficiency of all these methods has been that the
context  provided  by the local  tube  orientation  is  neglected  while
considering the  interaction  between  coarse-grained  units (Banavar et al.,
2002d) .  The novel
phase  discussed  here arises from the  addition  of  anisotropy  to the
well-studied  polymer  problem just as one obtains  rich liquid  crystal
behavior  on  studying  anisotropic  molecules.  A mapping  of the phase
behavior of tubes on varying the nature of  interactions,  the thickness
of the  tube,  the  length  of the  tube  and  temperature  might  yield
additional surprises.  \\

It is important to stress that our results are not at odds with or meant
as a substitute for the detailed and beautiful  work  involving the laws
of quantum  mechanics  and  biochemistry.  The virtue of our approach is
that it predicts a novel phase with selected types of structures and the
attendant   advantages.  It  is  then   necessary  to  complement   this
information with the principles of quantum chemistry to assess whether a
given biomolecule  would fit one of these  structures.  We do not invoke
hydrogen  bonds as Pauling did in his  prediction  of protein  secondary
motifs  (Pauling,  Corey and Branson, 1951; Pauling and Corey, 1951) and
indeed  not all the  structures  in the  marginally  compact  phase  are
compatible with hydrogen bond  placement.  What is remarkable,  however,
is that the lengths of the covalent and hydrogen  bonds and the rules of
quantum  chemistry  conspire  to  provide  a  perfect  fit to the  basic
structures  in this novel  phase.  One  cannot  but be amazed at how the
evolutionary  forces of Nature have shaped the molecules of life ranging
from the DNA molecule, which carries the genetic code and is efficiently
copied,  to  proteins,  the work  horses  of life,  whose  functionality
follows  from  their  form,  which, in turn, is a novel  phase of matter.
Protein  folds seem to be immutable -- they are not subject to
Darwinian   evolution  and  are  determined   from   geometrical
considerations,  as espoused by Plato (Denton and Marshall,  2001).
It is as if evolution  acts in the  theater  of life to  shape
sequences and  functionalities,  but does so within the fixed backdrop of
these Platonic folds.  \\

{\bf  Acknowledgments}  We are indebted to our collaborators  Alessandro
Flammini, Oscar Gonzalez, Trinh Hoang, John Maddocks, 
Cristian   Micheletti,   Flavio  Seno  and  especially Davide Marenduzzo
and Antonio   Trovato  for their
significant contributions to the work reported here.
We are grateful to Philip Anderson for valuable comments on a preliminary
version of the manuscript and George Rose for  many
stimulating  discussions.  This work was supported by Confinanziamento MURST,
INFM, NASA and the
Penn State MRSEC under NSF grant DMR-0080019.

\newpage \vspace{0.5in}


\newpage

FIGURE CAPTIONS

Figure 1:

Native state structure of the $B1$ domain (Protein data bank code: 1GB1)
of Protein G, a small protein produced by several Streptococcal species which
binds very tightly to Immunoglobulin.  
The domain shown has a length of 56
amino acids.   The structure contains an efficiently packed hydrophobic
core between a 4-stranded $\beta$-sheet (shown in blue) and a 4 turn 
$\alpha$-helix (shown in red).  
Strikingly, all protein  structures have helices, hairpins and sheets as
their building blocks. \\

Figure 2:

Space filling helix.

The geometry of this helix nicely illustrates 
the idea behind the 3-body
potential. Consider a tube in a compact helical conformation.  The smallest
value that the local radius of curvature of the helix can adopt equals the tube
thickness. Note that if the local radius of curvature
were any smaller than the tube thickness, the tube would self-intersect and
such configurations are not allowed.   Physically, a space-filling helix is
obtained when successive turns of the tube lie on top of each other. This
translates into the observation that the non-local radius associated
with three points, of which two are
close together and the third is alongside them 
in a neighboring turn, is also equal to the tube thickness.  Again, a radius
smaller than this value would lead to an intersection and is forbidden.
The  pitch  to  radius   ratio  of  this  helix  is  within  3 $\%$  of  the
corresponding value for $\alpha$-helices in globular proteins. \\

Figure 3:

Anti-parallel $\beta$-sheet, made of 4 strands, predicted by Linus Pauling.  
The local radius of curvature  of strands is greater  than that of helices,
but the non-local three-body radius associated with two neighboring  
$C_{\alpha}$ atoms
in a strand and a nearest  $C_{\alpha}$  atom in an adjoining  strand is
close to the local radius of curvature associated with a helix.  \\

Figure 4:

Sketch of the maximal number of contacts that a short, compact tube can
make as a function of $X$, the dimensionless
ratio of the tube thickness to the range
of the attractive interaction.  When $X$ is large compared to 1, one
obtains a swollen phase.  At the other extreme, when $X << 1$, one finds
a highly degenerate compact phase.  The twilight zone between these two 
phases occurs in the vicinity of $X \sim 1$ and is characterized by
marginally compact structures.  Typical tube conformations in each of
the phases are shown in the figure. \\

Figure 5:

Building blocks of biomolecules and ground state  structures  associated
with the marginally compact phase of a short tube. 
The top row shows some of the building blocks of biomolecules, while the
second row depicts the corresponding  structures  obtained for a tube in
the twilight zone.  (A1) is an  $\alpha$-helix  of a naturally  occuring
protein,   while  (A2)  and  (A3)  are  the  helices   obtained  in  our
calculations. (A2)  has a  regular  contact  map  whereas  (A3)  is a
distorted helix in which the distance between successive atoms along the
helical  axis is not  constant, but has period  $2$.  (B1) is a helix of
strands in the alkaline protease of pseudomonas aeruginosa, whereas (B2)
shows the corresponding  structure obtained in our computer simulations.
(C1) shows the  ``kissing''  hairpins of RNA and (C2) the  corresponding
conformation obtained in our simulations.  Finally (D1) and (D2) are two
instances of  quasi-planar  hairpins.  The first  structure  is from the
same protein as before (the alkaline protease of pseudomonas aeruginosa)
while the  second is a typical  conformation  found in our  simulations.
The sheet-like  structure  (D3) is obtained for a longer tube.  The tube
thickness  increases from left to right while the range of  interactions
is held fixed.  For more  details, see Banavar et al. (2002b).

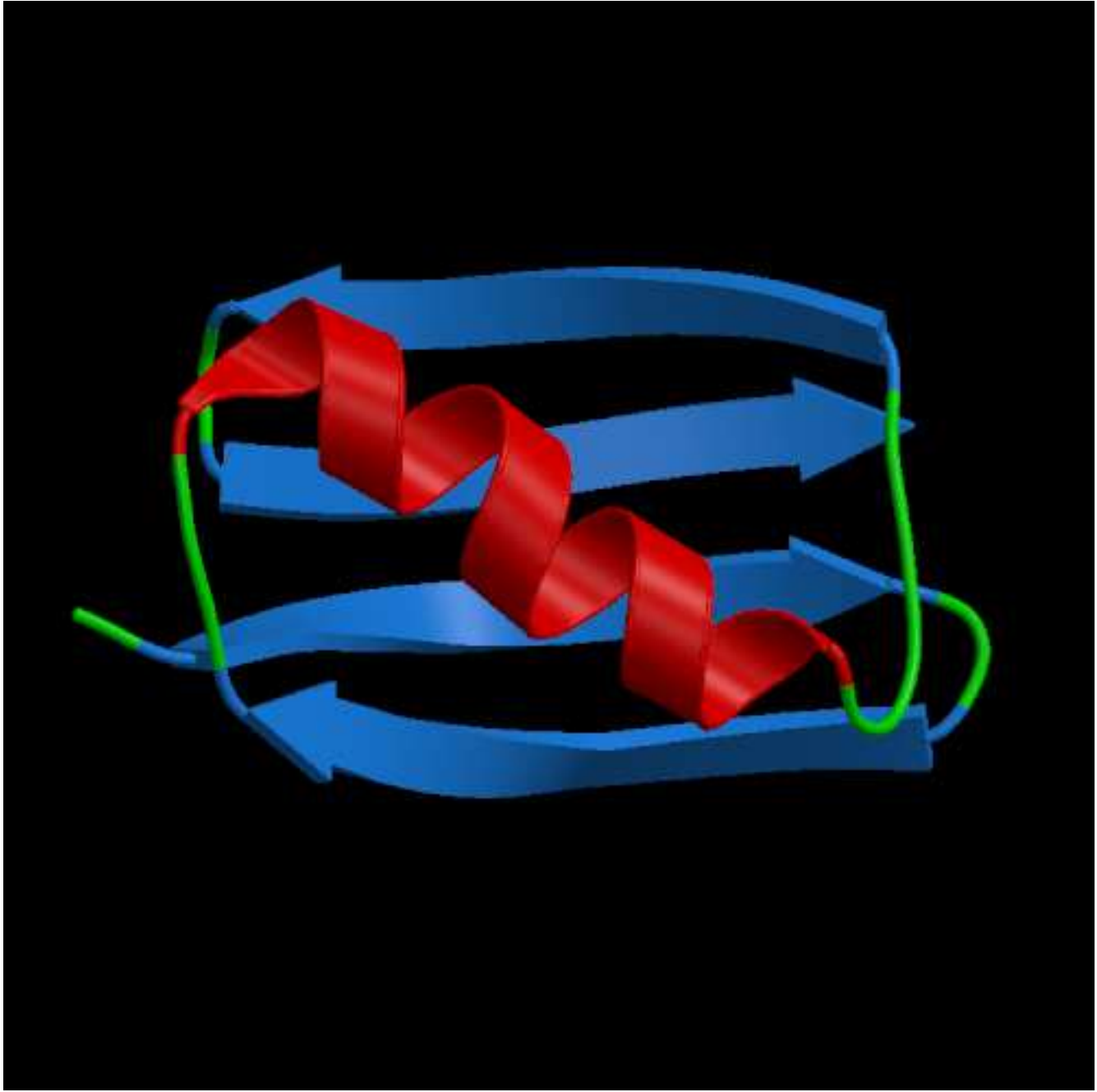
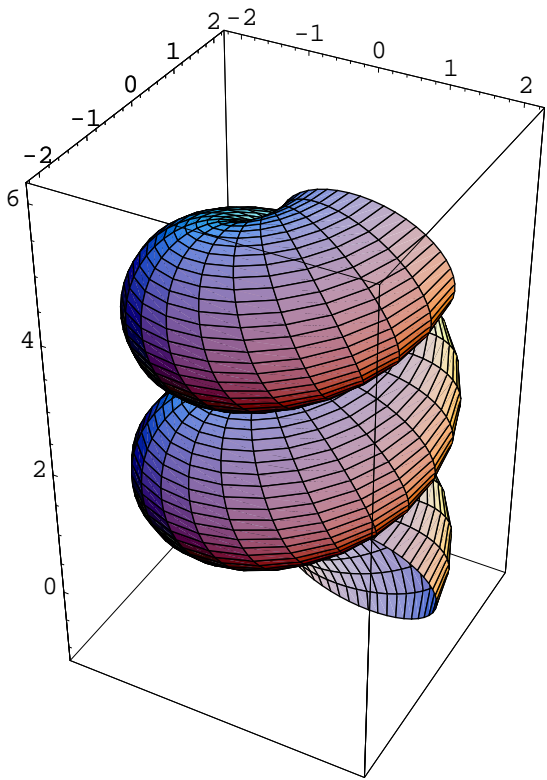
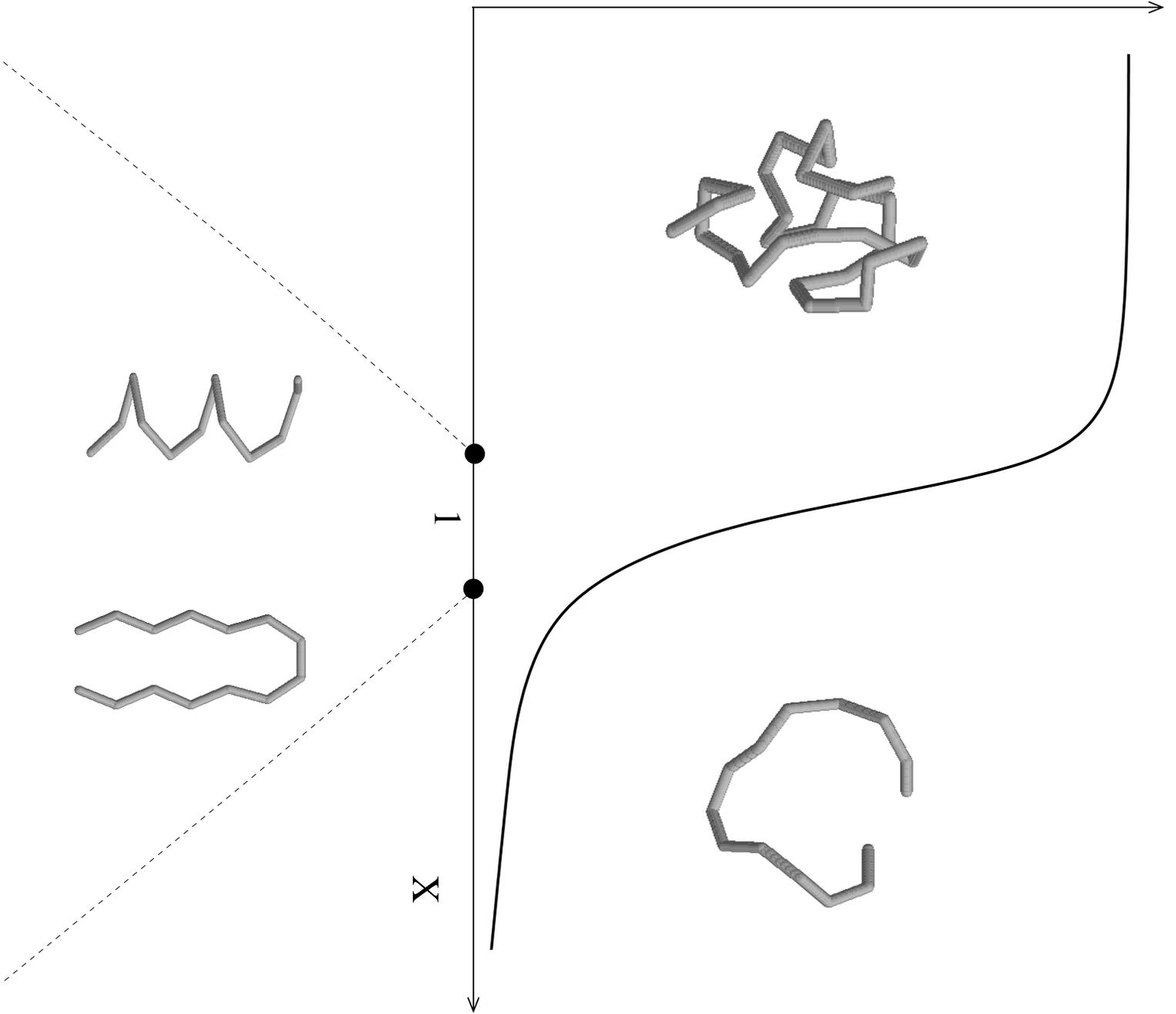
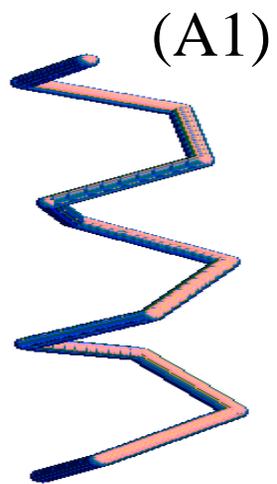
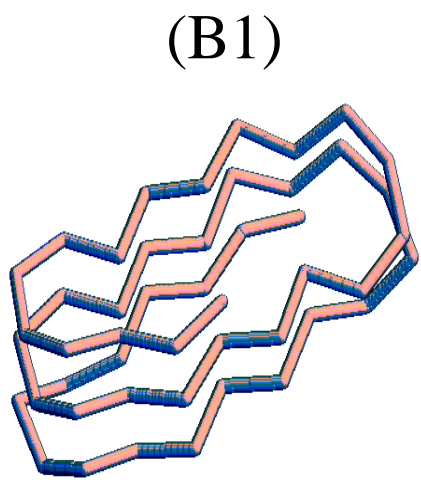
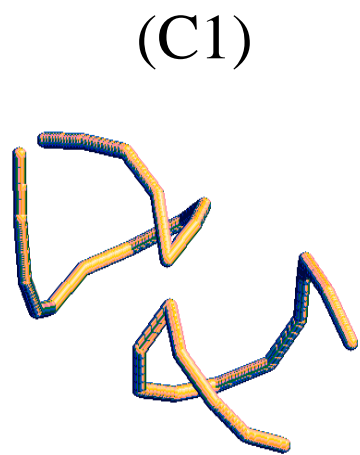
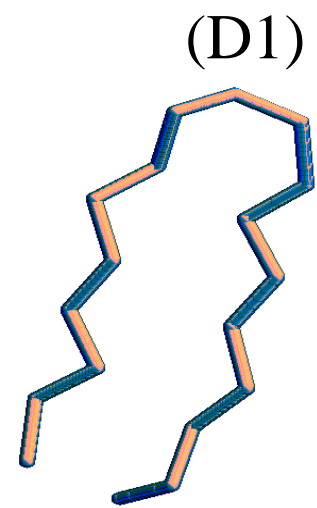
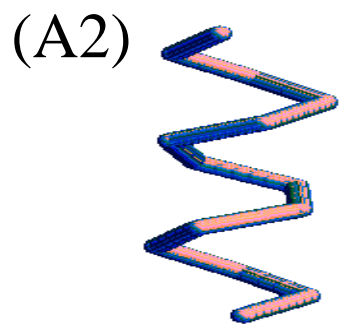
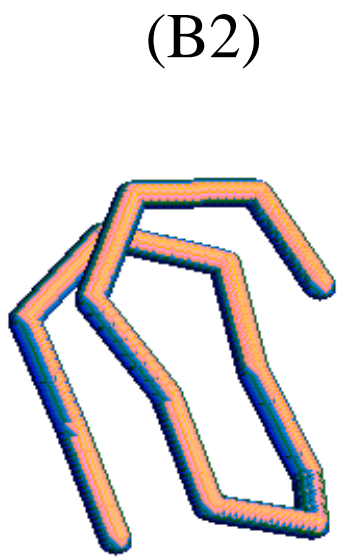
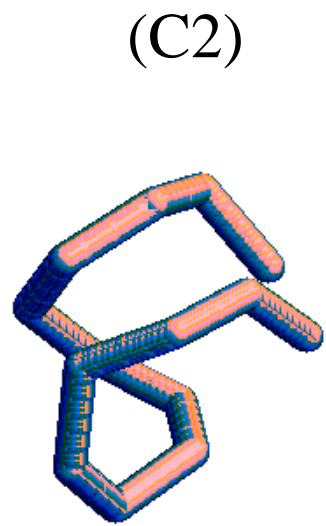
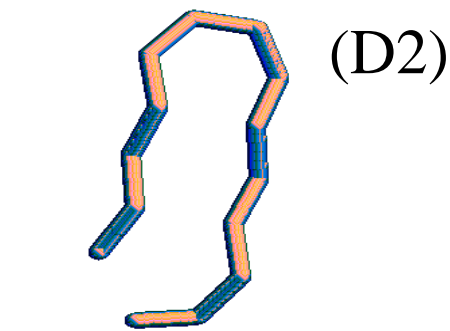
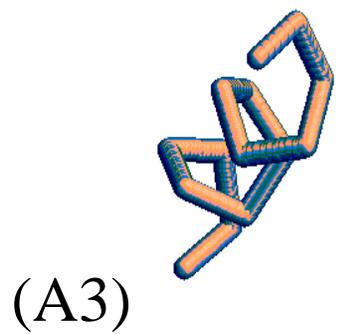
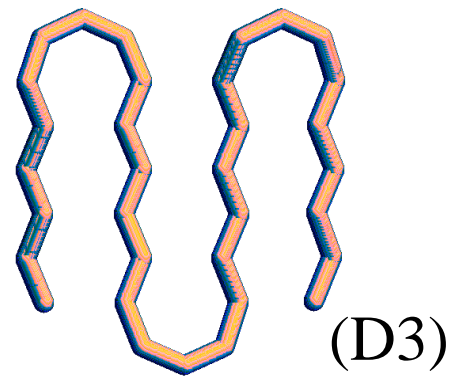
\end{document}